\documentclass[aps,prd,nofootinbib,longbibliography]{revtex4-2}
\usepackage{amsmath}
\usepackage{amssymb}
\usepackage{color}
\usepackage{graphicx}
\usepackage{rotating}
\usepackage{natbib}

\begin{document}

\title{A Many--Shapes framework for Galactic Cosmic Rays}

\author{Paolo Lipari}
\email{paolo.lipari@roma1.infn.it}

\affiliation{INFN, Sezione Roma ``Sapienza'',
 piazzale A.Moro 2, 00185 Roma, Italy}

\affiliation{
 IHEP, Key Laboratory of Particle Astrophysics,
 Chinese Academy of Sciences, Beijing, China}

\begin{abstract}
Thanks to recent measurements by space and
ground--level detectors, the energy spectra of cosmic ray protons
and helium nuclei
are now known with good precision over a broad energy range
extending for more than six orders of magnitude, from
a few GeV to approximately 10~PeV.
These measurements reveal that both spectra do not have a simple power--law form,
and exhibit several features.
The proton to helium ratio exhibits a very unexpected behavior: it decreases 
for rigidities $\lesssim 10^5$~GV, and increases for higher rigidities.
Several authors have interpreted the observed spectral features as
revealing the fact the CR spectra are formed by the contributions
of a few components (in some cases as few as two) produced by sources
that belong to distinct classes of astrophysical accelerators,
with sources in the same class producing spectra with the same
characteristic shape. Most of these authors also consider it
necessary to include the contribution of one (or perhaps more than one)
``local'' sources, recently active in the vicinity of the solar system.
However, this approach has several limitations, and
in this work we discuss an alternative scenario in which the cosmic ray sources,
even belonging to the same class, produce spectra with a broad
range of shapes. The power--law form of the total source spectrum
emerges as the average of these contributions.
In this framework observable spectral features can 
be produced by statistical fluctuations, and different
spectral slopes for protons and helium nuclei can be observed
even if one assumes that acceleration is rigidity--dependent.
\end{abstract}

\maketitle


\section{Introduction}
\label{sec:introduction}

Explaining the shape of the cosmic--ray (CR) spectra has been a major task in
the field for many decades.
As early as the 1940's it was understood that the CR spectrum 
extends to very high energies, and is, 
in first approximation, of power--law form.
In his groundbreaking paper 
``On the Origin of the Cosmic Radiation'' \cite{Fermi:1949ee} Enrico Fermi 
introduced a mechanism that ``yields naturally an inverse power law''
and that could perhaps reproduce the observed spectral slope ``of about 2.9''.
In this mechanism the CR particles, trapped in the Galaxy
by the magnetic fields that permeate interstellar space
(as had been just proposed by Alfvén \cite{alfven-book}),
are accelerated by multiple collisions
with moving magnetized clouds.

It is now understood that the Fermi mechanism, as originally proposed,
produces a CR spectrum that is too soft,
and it is generally accepted that the Galactic CR
are accelerated by an ensemble of discrete sources.
In this case the CR spectra observable at the Earth
are the source spectra, produced by the accelerators,
modified by propagation effects,
and therefore both the source spectra
and the distortions produced by propagation must have, at least in first
approximation, power--law form.

For several decades a ``standard model'' for the origin of the Galactic CR
has been under discussion: 
the ``supernova paradigm'' \cite{Ginzburg1964,Gaisser:2016uoy}.
This model has two main motivations. The first is related to the 
energy balance of the CR population: supernova explosions 
release in the interstellar space of the Milky Way
a very large power, of order $10^{42}$~erg/s 
in the form of kinetic energy of the ejecta.
If a fraction of order $\gtrsim 10$\% of this power is converted
in relativistic particles, this is sufficient to compensate for
the losses, mostly due to escape, and 
mantain a stationary CR population in the Galaxy. 
The second motivation is associated to the development of
Diffusive Shock Acceleration (DSA) theory.
In this theory CR are gradually accelerated by a sequence of interactions with
the shocks generated by violent astrophysical events (such as supernova explosions).
The fundamental acceleration mechanism is the same one 
originally proposed by Fermi, namely
acceleration of charged particles by the electromagnetic fields
carried by moving plasma, but in this case the shock geometry
allows to obtain harder spectra, in better agreement with the observations.

The simplest version of DSA, discussed in several textbooks \cite{Gaisser:2016uoy},
produces source spectra of power--law form with a slope $\Gamma \simeq 2 + 4/M^2$,
where $M$ is the Mach number of the shock in the unshocked medium frame,
so that, for strong shocks ($M \gg 1$) it results in a
``universal'' value $\Gamma \simeq 2$.
A source spectrum with such a slope is in fact very likely too hard,
because the best estimates of the propagation effects
obtain softenings that are too small to reproduce the shape of
the observed CR spectra. However, more realistic studies
of DSA, that include several additional effects, could
perhaps obtain softer spectra.

The power--law form of the CR spectra can only be valid in a limited energy range.
Deviations from a power--law form are expected, and observed, for
$E \lesssim $~few~GeV when the particles are non--relativistic.
In this low--energy range the CR spectra are also modified by time--dependent
solar modulations. In this work we will limit our discussion to
the range $E \gtrsim 50$~GeV, where solar modulations effects can be neglected.
On the other hand, the spectrum of the
particles accelerated in a source cannot extend to infinity,
and therefore the power--law must have a cutoff,
determined by either the containement of the particles
in the accelerators, or by the combination of
the acceleration rate and the lifetime of the source \cite{Hillas:1984ijl}.
A much discussed work by Lagage and Cesarsky \cite{Lagage:1983zz} estimated
that the maximum energy achieved in supernova remnants (SNRs)
is of order $E_{\rm max} \sim Z \times (10~{\rm TeV)}$
(with $Z$ the particle electric charge) suggesting that SNR sources cannot account
for the entire CR spectrum, and therefore that at least one additional
class of Galactic accelerators should exist. On the other hand, other studies,
(see for example Hillas \cite{Hillas:2005cs}) have discussed how,
taking into account other effects, in particular the non--linear
magnetic field amplification due to the CR population 
(an effect first discussed by Bell \cite{Bell:1978a}
and Bell and Lucek \cite{Bell:2001nlj}),
CR acceleration in SNRs could result in much higher maximum energies.

To include the effects of a maximum acceleration energy,
essentially all models constructed on the basis of the 
DSA mechanism predict that the spectrum of the accelerated particles
can be described by a power--law with an (approximately) exponential cutoff.
An important question is then
the prediction for the distribution of the cutoffs in different individual sources.
Naively one could expect that supernovae exploding in different conditions
and different environments should have significantly different cutoffs,
and similar considerations can be made for other type of accelerators.
However, it is also possible, and it is in fact commonly assumed,
that in all sources of the same class
is present a self--regulating mechanism that
produces approximately equal maximum energies
(see for example the discussion of Hillas
for supernovae \cite{Hillas:2005cs}).
In this case the effects of the cutoffs should be visible as features
in the observable CR spectra.

Recent, high precision measurements of the CR spectra of individual 
elements have revealed that they contain several features, that is deviations from a
simple power--law form.
This is inconsistent with the ``standard paradigm'' predictions
and understanding the origin of these spectral features is now
a fundamental problem for CR astrophysics.
Of particular interest are the observations of the
spectra of protons and helium nuclei because combining direct
and indirect measurements it is possible 
to study and compare the two spectra in a very broad energy range.

Several authors have published interpretation of the data
that are all based on the same fundamental idea:
the CR sources produce spectra that are well described by the
standard simple shape of a power--law with a cutoff,
but several different classes of astrophysical objects
contribute to CR production, and each class releases spectra with
a characteristic (class--dependent) shape.
The spectral features are then the result
of combining the contributions of different classes.
Many, in fact most, proponents of these ``multicomponent models''
note that it is also necessary
to include the contributions of one (or more)
``local'' sources, recently active in the vicinity of the solar system.

The aim of this work is to critically analyze the multicomponent models
and discuss their limitations. We will then propose an alternative
framework to describe the CR source spectra. In this framework
the individual accelerators that belong to the same class produce spectra 
that have many shapes, rather than a single, class--dependent shape.
The ensemble of all sources in one class produces a cumulative
spectrum that, in first approximation, has a power--law form.
However, this form
emerges as a statistical effect, resulting from the fact that
the power in CR emitted by softer sources is larger than the power
emitted by harder sources.

The work is organized as follows.
In the next section we perform a purely phenomenological study of the proton
and helium spectra in the energy range between 50~GeV and 10~PeV
and identify the features present in the two spectra.
In section~\ref{sec:multi-component-moodels} we discuss the
``multicomponent models'' adopted by several authors to explain
the origin of the observed spectral features.
In section~\ref{sec:many-shapes} we discuss the
alternative ``many--shapes'' framework
where the individual CR sources produce spectra
with a broad range of shapes, and the power--law form of the
cumulative spectra are  the result of  statistical averages.
The final section gives a summary and some tentative conclusions.

\section{The proton and helium energy spectra}
\label{sec:observations}
\begin{figure}[bt]
\begin{center}
\includegraphics[width=10.0cm]{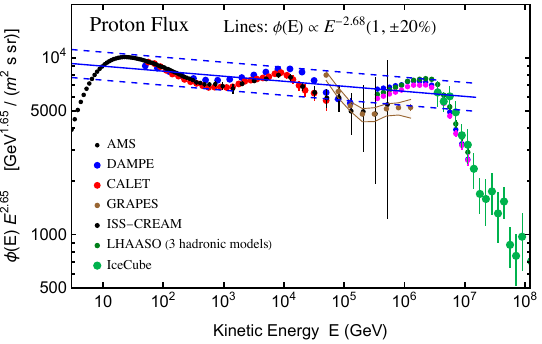}
\end{center}

\caption {\footnotesize
 Measurements of the flux of CR protons as
 a function of energy. The data are from 
 AMS02 \cite{AMS:2015tnn},
 DAMPE \cite{DAMPE:2025opn},
 CALET \cite{CALET:2022vro},
 GRAPES \cite{GRAPES-3:2024mhy},
 ISS--CREAM \cite{ISS-CREAM-Choi:2022aht},
 LHAASO \cite{LHAASO:2025byy}
 and IceCube \cite{IceCube:2019hmk}).
 To enhance the visibility of the spectral features
 the flux is shown in the form $\phi(E) \, E^{2.65}$ versus energy.
 The solid line is a simple power--law form ($\propto E^{-2.68}$)
 that is a best fit to the data in the energy interval between 10$^2$
 and $10^6$~GeV. The dashed lines have the same power--law form but
 normalizations that are $\pm 20$\% of the best fit.
 \label{fig:proton_flux1}
}
\end{figure}

\begin{figure}[bt]
\begin{center}
\includegraphics[width=10.0cm]{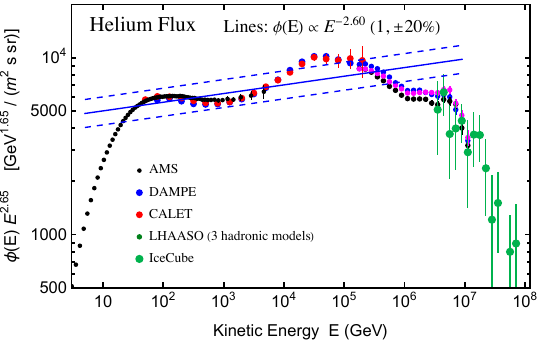}
\end{center}

\caption {\footnotesize
 Measurements of the flux of CR helium nuclei as
 a function of energy. The data are from 
 AMS02 \cite{AMS:2015azc},
 DAMPE \cite{DAMPE:2025opn},
 CALET \cite{CALET:2023nif},
 LHAASO \cite{LHAASO:2025mlf}
 and IceCube \cite{IceCube:2019hmk}).
 To enhance the visibility of the spectral features
 the flux is shown in the form $\phi(E) \, E^{2.65}$ versus energy.
 The solid line is a simple power--law form ($\propto E^{-2.60}$)
 that is a best fit to the data in the energy interval between 10$^2$
 and $10^6$~GeV. The dashed lines have the same power--law form but
 normalizations that are $\pm 20$\% of the best fit.
 \label{fig:helium_flux1}
}
\end{figure}

High--precision measurements of the proton and helium spectra 
have been obtained by the AMS magnetic detector \cite{AMS:2015tnn,AMS:2015azc},
by the calorimeters in near Earth orbit
DAMPE \cite{DAMPE:2025opn}, CALET \cite{CALET:2022vro,CALET:2023nif} and 
ISS--CREAM \cite{ISS-CREAM-Choi:2022aht}, and by Extensive Air Shower (EAS)
detectors at ground level: GRAPES--3 \cite{GRAPES-3:2024mhy},
LHAASO \cite{LHAASO:2025byy,LHAASO:2025mlf} and IceCube \cite{IceCube:2019hmk}.
These measurements are shown,
plotted as a function of energy,
in Figures~\ref{fig:proton_flux1} and~\ref{fig:helium_flux1}.
In these figures,
the fluxes are shown
multiplied by the energy--dependent factor $E^{2.65}$
to enhance the visibility of possible features.
There are some differences between the measurements
but overall there is remarkably good agreement.
It is remarkable that the measurements taken in adjacent energy ranges with 
direct methods and by air shower observations connect smoothly.
Therefore, a consistent picture of the structure of the two spectra emerges,
and several deviations from a simple power--law form are clearly visible
and identifiable.

Additional measurements of the CR spectra have been obtained by EAS telescopes
also at higher energies
(see for example \cite{TelescopeArray:2015dcv,TelescopeArray:2018bya,PierreAuger:2025eun}).
In this very high energy range it is difficult to separate
the spectra of different elements due to large systematic uncertainties.
A discussion of these results is deferred to future work.

The largest deviations from a simple power--law
are present a $E \simeq 3$~PeV for protons and $E \simeq 6$~PeV for helium nuclei,
where the spectra have marked softenings that can be
associated to the well known spectral break
first observed in the all--particle CR spectrum
already in the 1950's \cite{kulikov:1959} and known as the ``knee''.
Other less prominent features are visible at lower energies.
As discussed below, one can identify three additional features
in the proton spectrum and four in the helium spectrum.

It is interesting to note that in the broad energy range
between $E \simeq 50$~GeV and the ``knee'' 
the deviations of the two spectra from a simple power--form are small.
In this range a power--law with slope $\Gamma_p^* \simeq 2.68$ describes the
proton spectrum with deviations $ \lesssim \pm 20$\%. 
A little harder power--law with slope $\Gamma_{\rm He}^*\simeq 2.61$
describes the helium spectrum with deviations of the same order.
To visualize these results the power--law approximations of the
two spectra, and the $\pm 20\%$ bands are also plotted in 
Figures~\ref{fig:proton_flux1} and~\ref{fig:helium_flux1}.
In these figures the exponent 2.65 of the energy--dependent factor
that multiplies the spectra was chosen because it is close to the 
slopes of the two power--law approximations, and therefore
maximizes the visibility of the features.

For a more detailed phenomenological description of the spectra
we have fitted the data using the multiple--break expression \cite{Lipari:2017jou}:
\begin{equation}
 \phi(E) = \phi_0 \, \left (\frac{E}{E_0} \right )^{-\Gamma_0}
 \; \prod_{j=1}^{N_b} \left [ 1 - \left ( \frac{E}{E_{b,j}}\right )^{1/w_j}
 \right ]^{-(\Gamma_j- \Gamma_{j-1}) \, w_j} ~.
 \label{eq:multi_break}
\end{equation}
This expression contains $N_b$ features (or breaks) where the
spectral index undergoes a step,
changing from $\Gamma_{j-1}$ to $\Gamma_j$.
A step is symmetric in $\ln E$, centered
at $\ln E_{b,j}$ (with $E_{b,j}$ the break energy of feature $j$)
and has width $\Delta \ln E \simeq \pm w_j$.
Therefore the fit, 
including the low energy slope $\Gamma_0$ (for $E \ll E_{b,1}$) and
the normalization parameter $\phi_0$,
depends on $3 N_b + 2$ parameters ($E_0$ is an arbitrary reference energy).

The energy dependent slope (or spectral index) has the form:
\begin{equation}
 \Gamma(E) = - \frac{d \ln \phi(E)} {d \ln E}
 = \Gamma_0 + \sum_{j=1}^{N_b} \frac{(\Gamma_j - \Gamma_{j-1})}{2} \;
 \left [ 1 + \tanh \left ( \frac{ \ln E - \ln E_{b,j}} {2 \, w_j} \right ) \right ] ~.
 \label{eq:slope}
\end{equation}
If the width $w_j$ are sufficiently small
$\Gamma_j$ is the slope of the spectrum
in the range $\ln E_{b, j} \lesssim \ln E \lesssim E_{b,j+1}$.
The spectral curvature is
\begin{equation}
 \beta(E) = \frac{1}{2} \frac{d\Gamma(E)}{d \ln E} =
 \frac{1}{8} \; \sum_j \frac{(\Gamma_j - \Gamma_{j-1})}{w_j} \; \left [\cosh 
 \left ( \frac{ \ln E - \ln E_{b,j}} {2 \, w_j} \right ) \right ]^{-2}~
\label{eq:curvature}
\end{equation}
and is significantly different from zero only in the regions 
$|\ln E - \ln E_{b,j}| \lesssim w_j$.

Fitting the observations is is possible to unambiguosly identify, in the energy range between
50~GeV and 10~PeV, four features in the proton spectrum
and five features in the helium spectrum.

\vspace{0.25cm}
The four features in the proton spectrum, 
in order of increasing energy are: 
\begin{itemize}
\item [(p1)] A hardening centered at $E \simeq 0.5$~TeV.
\item [(p2)] A softening centered at $E \simeq 10$~TeV.
\item [(p3)] A second hardening centered at $E \simeq 100$~TeV.
\item [(p4)] A marked softening at $E \simeq 3$~PeV that corresponds to
 the well known CR knee.
\end{itemize}
Features (p1), (p2) and (p4) have been extensively discussed in the literature.
The existence of a hardening below the knee was first proposed
in \cite{Lipari:2019jmk} as necessary to connect the direct and Air--Shower
measurements of the spectrum, and was then directly measured by the GRAPES--3
detector \cite{GRAPES-3:2024mhy}. Such a hardening is clearly necessary
to connect the measurements of CALET and DAMPE with those obtained by LHAASO.

\vspace{0.25cm}
The five features in the helium spectrum (described in terms of rigidity) are:
\begin{itemize}
\item [(h1)] A hardening centered at rigidity $\mathcal{R} \simeq 0.5$~TV.
\item [(h2)] A softening centered at rigidity $\mathcal{R} \simeq 10$~TV.
\item [(h3)] An additional softening around 
 $\mathcal{R} \simeq 100$~TV.
\item [(h4)] A hardening around $\mathcal{R} \simeq 500$~TV.
\item [(h5)] A softening, that can naturally associated to the knee at rigidity $\mathcal{R} \simeq 3$~PV.
\end{itemize}

\begin{figure}[bt]
\begin{center}
\includegraphics[width=15cm]{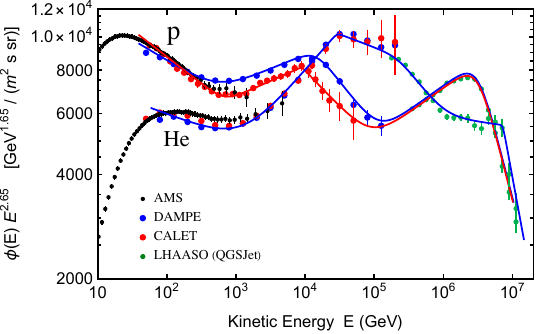}
\end{center}

\caption {\footnotesize
 Fluxes of protons and helium nuclei as a function of kinetic energy.
 The data points are from
 AMS02 \cite{AMS:2015tnn,AMS:2015azc},
 DAMPE \cite{DAMPE:2025opn},
 CALET \cite{CALET:2022vro,CALET:2023nif}
 and LHAASO (QGSJet model) \cite{LHAASO:2025byy,LHAASO:2025mlf}.
 The lines are
 fits to the data using
 the multi--break
 expression of Eq.(\ref{eq:multi_break}).
 \label{fig:prot_helium_flux_energy}
}
\end{figure}

The fits to the proton and helium spectra
are shown together with the data in Fig.~\ref{fig:prot_helium_flux_energy}.
Combining data of different detectors that are not in
perfect agreement is not an easy task, and for this reason we 
have performed two different fits
for the data combinations (AMS, DAMPE, LHAASO) and (AMS, CALET, LHAASO),
in both cases we have excluded the three lowest data points of the LHAASO
proton spectrum because they have large systematic errors.
Both fits are shown in the figure, and the best fits parameters are shown
in Tables~\ref{tab:par1} and~\ref{tab:par2}.
The differences between the two fits can be considered as an estimate of the systematic
uncertainties and have no impact of the following discussion.

The best way to compare the proton and helium spectra
is to study them as a function of rigidity, because the simplest acceleration models
predict source spectra of the same shape and if absorption is negligible,
also the effects of propagation are also rigidity dependent.
It is therefore unexpected that the shapes in rigidity
of the proton and helium spectra are not identical.
Three of the features in the helium spectrum (h1), (h2) and (h5)
correspond to the proton features (p1), (p2) and (p4) and are centered, in good approximation,
at the same rigidites (0.5~TV, 10~TV and 3~PV),
however in the range $\mathcal{R} \in [30~{\rm TV}, 1000~{\rm TV}]$
the shapes of two spectra differ. The proton spectrum contains one clear hardening
centered at $\mathcal{R}\simeq 100$~TV, while at this rigidity the helium spectrum
has the opposite behavior and suffers an additional softening.
The helium spectrum does contain a significant hardening but at a rigidity
of order 500~TV, five times higher than for protons.

\begin{figure}[bt]
\begin{center}
\includegraphics[width=12.0cm]{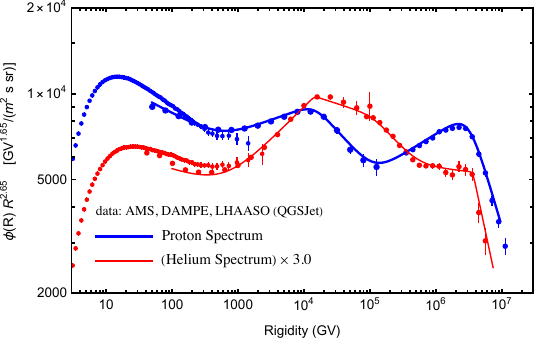}

\vspace{0.9cm}
\includegraphics[width=12.0cm]{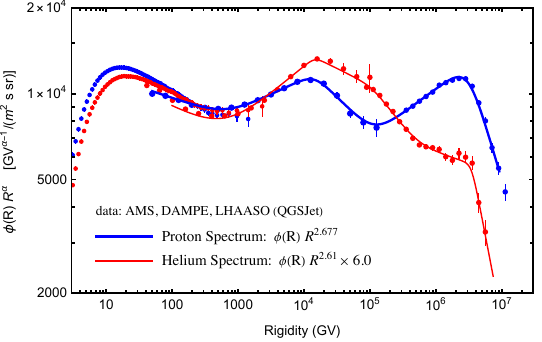}
\end{center}

\caption {\footnotesize
Spectra of protons and helium nuclei as a function of rigidity.
In the top panel, to allow a better visualization of the shapes,
the two spectra are plotted in the form
$\phi(\mathcal{R}) \, R^{2.65}$ versus $\mathcal{R}$, with the helium flux also
multiplied by a factor of three.
In the bottom panel the two spectra are shown multiplied by different rigidity
dependent factors. The proton spectrn is plotted 
in the form $\phi(\mathcal{R}) \, \mathcal{R}^{2.677}$ versus $\mathcal{R}$,
while the helium spectrun is plotted in the form
$6 \phi(\mathcal{R}) \, R^{2.610}$ versus $\mathcal{R}$.
The data points are from
AMS02 \cite{AMS:2015tnn,AMS:2015azc},
DAMPE \cite{DAMPE:2025opn},
and LHAASO (QGSJet model) \cite{LHAASO:2025byy,LHAASO:2025mlf}.
The lines are fits to the data using the multi--break
expression of Eq.(\ref{eq:multi_break}).
\label{fig:spectrum_rigidity}
 }
\end{figure}

To compare the shapes in rigidity of the spectra one can inspect Fig.~\ref{fig:spectrum_rigidity}
where the two spectra are plotted together as a function of rigidity.
In the top panel both spectra are shown in the form
$\phi (\mathcal{R}) \times \mathcal{R}^{2.65}$, with the helium flux multiplied
by an additional constant factor of three.
It is apparent that the two spectra have significantly different shapes.

In the bottom panel of Fig.~\ref{fig:spectrum_rigidity}
the two spectra are shown multiplied by two different rigidity dependent factors:
$\mathcal{R}^{2.677}$ for protons and $6 \times \mathcal{R}^{2.61}$ for helium.
This choice is motivated by the fact that plotted with these
different weights the two spectra are approximately equal in the
rigidity range $\mathcal{R}\lesssim$~few~TV. This result was first discussed by the 
AMS Collaboration in \cite{AMS:2015azc}, where it is noted 
that in the rigidity range from 40 to 3000~GV (where both spectra undergo a hardening)
the $p$/He ratio is consistent with a simple power--law with a slope $0.077 \pm 0.002\pm 0.007$.

Some authors \cite{Malkov:2011gb,Hanusch:2018bsk} have suggested
that such a difference in spectral shape,
while not anticipated, is not only consistent
with the standard Diffuse Shock Acceleration (DSA) mechanism
but is in fact a signature of this mechanism, that according to the authors
produces power--law of different slopes for the two particle types.

\begin{figure}[bt]
\clearpage
 
\begin{center}
\includegraphics[width=12cm]{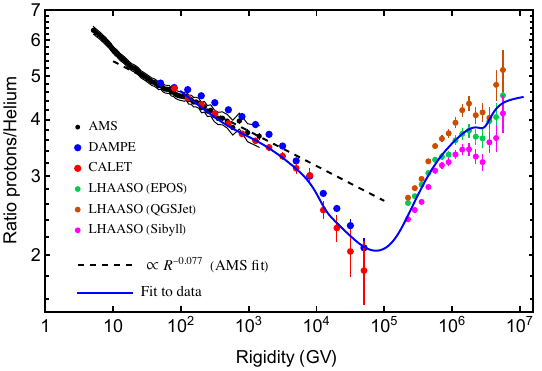}
\end{center}

\caption {\footnotesize
 Ratio of the fluxes of protons and helium nuclei as a function of
 rigidity. 
 The data points are from
 AMS02 \cite{AMS:2015tnn,AMS:2015azc},
 DAMPE \cite{DAMPE:2025opn},
 CALET \cite{CALET:2022vro,CALET:2023nif} and
 LHAASO \cite{LHAASO:2025byy,LHAASO:2025mlf}.
 The solid line is the ratio of fits to the proton and helium
 spectra shown in Fig.~\ref{fig:spectrum_rigidity}.
 The three data sets for LHAASO have  been calculated with
 three hadronic interaction models:
 QGSJet \cite{Ostapchenko:2013pia},
 EPOS \cite{Pierog:2013ria}
 and Sibyll \cite{Riehn:2019jet}.
 \label{fig:ratio_rigidity}
}
\end{figure}

However, the extension of the observations to higher rigidities shows that
the relation between the proton and helium spectra has a much more
complex rigidity dependence than the simple power--law observed below few TV.
This can also be clearly seen in Fig.~\ref{fig:ratio_rigidity}
that shows some measurements of the $p$/He ratio 
plotted as as a function of rigidity.
For low rigidities the ratio follows the power--law behavior
discussed in the AMS publication \cite{AMS:2015tnn},
however for $\mathcal{R} \gtrsim 3$~TV
(above the range of the AMS observations)
the ratio decreases more rapidly than the extrapolation of the AMS fit.
Even more surprising are the results 
obtained by LHAASO at higher rigidities that in the range $\mathcal{R} \gtrsim 10^5$~GV
observes a $p$/He ratio with the opposite behavior, 
increasing with rigidity.

\begin{figure}[bt]
\begin{center}
\includegraphics[width=12cm]{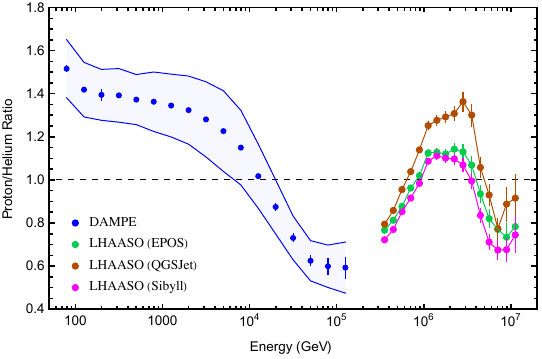}
\end{center}

\caption {\footnotesize
 Ratio of the fluxes of protons and helium nuclei as a function of energy.
 The data points are from DAMPE \cite{DAMPE:2025opn} and
 LHAASO \cite{LHAASO:2025byy,LHAASO:2025mlf}. One can note
 three crossing energies at approximately 10~TeV, 500~TeV abd 3.5~PeV
 where the $p$/He ratio is unity.
 \label{fig:ratio_pHe_energy}
}
\end{figure}

The $p$/He ratio can be also studied as a function of energy,
and Fig.\ref{fig:ratio_pHe_energy} shows the measurements of the
$p$/He ratio by DAMPE and LHAASO plotted as a function of energy
in the range between 100~GeV and 10~PeV.
It is remarkable that protons and helium nuclei
replace each other three times as the largest CR component at a fixed
value of the particle energy.
The three crossings, where the $p$/He ratio is unity, are at 
$E \simeq 10$~TeV, 500~TeV and 3.5~PeV.
At low energies the helium flux is smaller than the proton one,
but it is also harder, so that the 
$p$/He ratio, that is initially much larger than one, decreases gradually,
becoming unity at the first crossing energy ($E \simeq 10$~TeV).
The $p$/He ratio continues to decrease with energy reaching a minimum value of order 0.6
at $E \approx 100$~TeV. At this energy 
the proton spectrum undergoes a hardening while the helium flux
softens, so that the energy dependence of the ratio changes it behavior
and starts to increase. The ratio becomes again unity at a second crossing point
at an energy of order $E\simeq 500$~TeV, and continues to grow.
This growth stops at $E \approx 2$~PeV when the knee softening starts to develop in the proton
spectrum while the helium flux keeps its power--law form,
and the $p$/He ratio returns to be decreasing, 
and one observes a third crossing point at $E \approx 3.5$~PeV,
where helium nuclei become again the largest CR component.
The energy dependence of the $p$/He ratio could have additional structure
after the knee of the helium component, but the measurement errors
are too large to reach solid conclusions.

\begin{figure}[bt]
\begin{center}
\includegraphics[width=15cm]{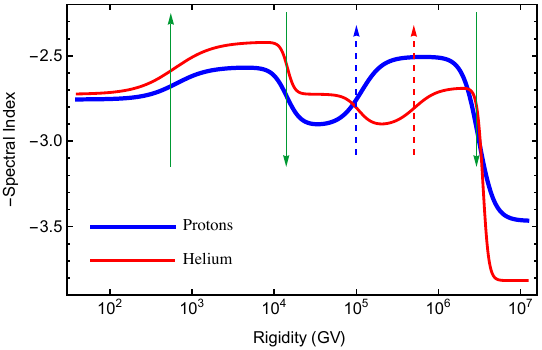}
\end{center}

\caption {\footnotesize
 Spectral index of fits to the measurements of the spectra
 of CR protons and helium nuclei with the multi--break
 expression of Eq.(\ref{eq:multi_break}).
 The spectral index is shown in the form $-\Gamma(E)$ versus energy.
 The arrows show the positions of the center of the most prominent features
 in the spectra where the spectral index undergoes a step.
 Three of these features (an hardening at $\mathcal{R} \simeq 0.5$~TeV,
 a softening at $\mathcal{R} \simeq 14$~TeV and the ``knee'' at
 $\mathcal{R} \simeq 3000$~TeV)
 are at approximately the same rigidity in both the proton and
 helium spectra. A second hardening appears at different
 rigidities for the two elements.
 ($\mathcal{R} \simeq 100$~TeV for protons and
 $\mathcal{R} \simeq 500$~TeV for helium).
 There is also evidence that the fit to the helium flux
 contains an additional softening feature at a rigidity of $\sim 100$~TeV.
 \label{fig:slope_rigidity}
}
\end{figure}

The information on the shapes of the $p$ and He spectra is also encoded 
in the rigidity dependence of their slopes.
Figure~\ref{fig:slope_rigidity} shows the slopes of the two spectra
obtained from fits to the data plotted as a function of rigidity
(with a minus sign, so that the harder spectrum is above the softer one).
One can see that there are rigidity intervals
where the slopes are approximately constant, and the spectra are well
approximated by a power--law shape, and narrower intervals
where the slope changes that correspond to the spectral features.
Vertical arrows indicate the positions of the centers of the spectral features.
It is again apparent that the proton spectrum contains four features
(two hardenings and two softenings) while in the helium spectrum one can recognize five features.

The most striking aspect of this study is that
the helium spectrum is harder that the proton one 
for $\mathcal{R} \lesssim 100$~TV, and softer at higher rigidities.
The intriguing result noted by the AMS collaboration 
that the difference $\Gamma_p (\mathcal{R}) - \Gamma_{\rm He} (\mathcal{R})$
is a constant is only valid for low rigidities ($\mathcal {R} \lesssim 3$~TV).

In the following we will first discuss interpretations
of these observations based on the multicomponent framework,
and then discuss a possible alternative.

\section{Multicomponent models of the cosmic ray spectra}
\label{sec:multi-component-moodels} 

\subsection{The formation of the cosmic--ray spectra}
It is generally accepted that the formation of the CR fluxes
can be analyzed into two, reasonably well separated parts:
acceleration and propagation.
Assuming that the energy of the CR particles remain constant
during propagation, the flux 
$\phi_Z (E, \vec{x}_{\rm obs},t_{\rm obs}$) of nuclei of
electric charge $Z$ and energy $E$ observable at the point $\vec{x}_{\rm obs}$
at time $t_{\rm obs}$ can be expressed as the convolution: 
\begin{equation}
 \phi_Z (E,\vec{x}_{\rm obs}, t_{\rm obs}) =
 \frac{c}{4 \pi} \; \int d^3x_s \; \int dt_s ~Q_a (E, \vec{x}_s, t_s) \;
 \mathcal{P}_Z (\vec{x}_{\rm obs}, \vec{x}_{s}, t_{\rm obs}, t_{s}; E)~. 
\label{eq:prop_general}
\end{equation}
In this equation $Q_Z (E, \vec{x}_s, t_s)$
(with dimension $E^{-1} \, L^{-3}\, T^{-1}$) is the rate of
injection in interstellar space of nuclei of charge $Z$
from the point $\vec{x}_s$ at the time $t_s$.
The propagation function $\mathcal{P}_Z (\vec{x}_{\rm obs}, \vec{x}_{s}, t_{\rm obs}, t_{s}; E)$
(with dimension $L^{-3}$) gives the probability 
of finding at the time $t_{\rm obs}$ and in the vicinity of
the point $\vec{x}_{\rm obs}$ 
a CR particle injected in interstellar space with energy $E$ at time $t_s$
from the point $\vec{x}_s$. If absorption is negligible,
then the propagation function is a function of the rigidity
$\mathcal{R} = E\beta/(Z e)$ for all nuclei,
and if the Galaxy can be considered in a
stationary state it depends only on the time difference $t = t_{\rm obs} - t_s$.
The factor $c/(4 \pi)$ in Eq.~(\ref{eq:prop_general}) converts 
a particle density in a flux assuming that the particles are ultrarelativistic,
and have an isotropic angular distribution.

It is generally accepted that primary CR particles
(such as protons and helium nuclei) are accelerated in 
discrete sources that are transients and well localized in space.
Accordingly, the CR injection rate can be written as the sum:
\begin{equation}
 Q_Z (E, \vec{x}_s, t_s) = \sum_j q_{j,Z} (E) \;
 \delta[\vec{x}_s - \vec{x}_j] \;
 \delta[t_s - t_j] ~,
\label{eq:qz}
\end{equation}
where $q_{j,Z} (E)$ is the spectrum of nuclei of charge $Z$,
injected in space from the source $j$ (at position $\vec{x}_j$ and time $t_j$).

A large body of the literature on CR acceleration,
has been based on the assumption that the CR accelerators are all of the same type
(presumably supernova explosions) and produce
rigidity--dependent spectra with a unique characteristic shape:
a power--law with a universal slope and a high--energy cutoff
(associated to the maximum acceleration energy in the sources)
proportional to the electric charge of the particles:
\begin{equation}
 q_{j,Z} (E) = q^{(0)}_{j,Z} \; \left (\frac{E}{E_0} \right)^{-\Gamma}
 ~\exp \left [-\frac{E}{Z \, E_{{\rm cut},p}} \right ] 
\label{eq:q-standard}
\end{equation}
where $E_0$ is an arbitrary reference energy and the
quantities $q^{(0)}_{j,Z}$ are constants that depend on the source
and the particle type.

The motivations for the power--law form of the source spectra
(with a universal value of the slope) 
and for the rigidity dependence of the high--energy cutoff
were based on DSA theory. In fact, DSA theory does not require
the cutoffs to be equal in different sources. 
The assumption of equal (or approximately equal) cutoffs
has a phenomenological motivation, as it allows
to interpret the ``knee'' softening
as the signature of this universal maximum acceleration energy.

To relate a model the accelerators to the observations it is
obviously also necessary to have a model for CR propagation.
To construct such a model two approximations are often introduced:
\begin{itemize}
\item [(i)] The CR injection can be described by its
 (time--independent) time average $\overline{Q}_Z (\mathcal{R}, \vec{x}_s)$,
  and its rigidity and spatial dependence can be factorized.
  Expressing the CR injection as a function of rigidity one has:
 \begin{equation}
 \overline{Q}_Z (\mathcal{R}, \vec{x}_s) =
 \frac{1}{(t_{\rm max} - t_{\rm min})} ~\int_{t_{\rm min}}^{t_{\rm max}} dt
 ~Q_Z(\mathcal{R}, \vec{x}_s,t) = \overline{Q}_Z^{\rm tot} (\mathcal{R}) \; F_Z ( \vec{x}_s)
 \end{equation}
 where $\overline{Q}_Z^{\rm tot} (\mathcal{R})$ is the average CR injection rate
 from the entire Galaxy, and 
 $F_Z(\vec{x}_s)$ is the normalized spatial distribution
 for the injection in interstellar space of nuclei with electric charge $Z$.
\item [(ii)] Propagation can be described by diffusion with
 a diffusion coefficient $D(\mathcal{R})$ that has the same rigidity dependence
 in all the Galactic confinement volume.
\end{itemize}
The observable flux takes then the formL:
\begin{equation}
 \phi_Z (\mathcal{R}) =
 \frac{c}{4 \, \pi} \;
 \frac{\overline{Q}_Z^{\rm tot} (\mathcal{R}) } {D(\mathcal{R})} ~ K_Z 
~.
 \label{eq:flux_factorized}
\end{equation}
and is proportional to the product of the total Milky Way 
CR injection spectrum and the inverse of the diffusion coefficient,
with $K_Z$ a constant (with dimensions $L^{-1}$)
that depends on the spatial distribution of the sources
(the $Z$ subscript allows for different spatial distributions for
different nuclei).

The factorization of the rigidity and
spatial distribution of the CR injection is automatically obtained
assuming that all CR sources produce spectra of the same shape.
On the other hand, the use of a time averaged
injection to compute the observable spectra can in some cases
give incorrect results if these spectra receive significant contributions
from ``local'' sources, that is sources active in the very recent past
with a position very close to the observation point.
The contribution of a ``local''
source depends on its age and position and its inclusion
requires a special treatment.

According to Eq.(\ref{eq:flux_factorized}), the shapes of the
observable spectra are determined by the rigidity dependences of the
two factors (injection and diffusion coefficient),
and the slope of the observable flux $\Gamma_{\phi,Z} (\mathcal{R})$ is the sum:
\begin{equation}
 \Gamma_{\phi,Z} (\mathcal{R}) =
- \frac{d\ln \phi_{Z}(\mathcal{R})}{d\ln \mathcal{R}}
 =
 - \frac{d\ln \overline{Q}_{Z}^{\rm tot} (\mathcal{R})}{d\ln \mathcal{R}}
 + \frac{d\ln D(\mathcal{R})}{d\ln \mathcal{R}}
 = \Gamma_{Q,Z} (\mathcal{R}) + \delta (\mathcal{R}) ~.
\end{equation}

Generally accepted models for the Galactic cosmic rays assumed
that the diffusion coefficient had a simple power--law form
($D(\mathcal{R}) \propto \mathcal{R}^\delta$) with a constant
exponent $\delta$, while the injection had the form of
Eq.(\ref{eq:q-standard}). This lead to the prediction
that the CR spectra, for all primary nuclei,
should be simple power--laws with equal slopes
and an exponential cutoff at the same rigidity.
These expectations are not supported by the observations
that revealed the existence of several spectral features.
Multicomponent models were developed to explain these observations.

\subsection{Multicomponent models}
\label{sec:multi-components}
The multicomponent approach was first introduced by
Gaisser, Stanev and Tilav (GST2013) in \cite{Gaisser:2013bla}.
The authors of this paper proposed a phenomenological, data motivated model
for the CR spectra with the aim of describing its very high energy features:
the ``second knee'' at $E \simeq 10^{17}$~eV and the
``ankle'' $E \simeq 5 \times 10^{18}$~eV \cite{Blumer2009Review,Apel2012Spectrum}.

In the model the CR spectrum for the element $Z$ is formed by the sum of the contributions
from a few (three or four) source classes that produce spectra
that have a power--law form with an exponential cutoff:
\begin{equation}
 \phi_Z (E)
 = \sum_{C \in {\rm classes}} \phi_{Z,C}(E) 
 = \sum_{C \in {\rm classes}} K_{Z,C} \; \left (\frac{E}{E_0} \right )^{-\Gamma_{Z,C}} 
 \exp \left [ - \frac{E} {Z \, E^{{\rm cut}}_{p,C}} \right ]
\end{equation}
The model implicitely assumes that the effects of propagation 
generate a featureless (power--law form) distortion of the source spectra,
and the differences in the spectral shape of the fluxes are 
attributed to the properties of the accelerators.

The GST2013 approach is a straightforward generalization of what
had been the commonly accepted model for the formation of the
CR spectra described in the previous subsection.
Each class of sources produces spectra with the form
of Eq.(\ref{eq:q-standard}) a power--law with a high--energy
exponential cutoff. The high--energy cutoffs for different
nuclei in the same source class are proportional to the electric charge $Z$
(in other words they are cutoffs at a fixed rigidity),
so that each class of sources forms a ``Peters cycle'' \cite{Peters:1961mxb}.
The model however considers the slopes $\Gamma_{Z,C}$ as free parameters
that can be fitted to the data, and allows for the spectra
of different elements accelerated in the same sources to have
different slopes. This is inconsistent with the idea that
acceleration is rigidity independent, but it appears
necessary to describe the data.

\subsection{Multicomponent models and the proton and helium spectra}
Several authors have published papers interpreting the proton and helium
spectra in a multicomponent framework. These papers 
discuss  the identification of the number and properties of the
classes of sources that make significant contributions.
Some of these authors also argue that, to accurately describe the data,
it is necessary to introduce the contribution of one
(or possibly two) ``local'' sources.

As discussed above in Section~\ref{sec:observations},
the proton and helium fluxes decrease very rapidly with energy (or rigidity).
In the range studied in this work, the two spectra can be approximated
by simple power--laws with slopes of approximately 2.68 and 2.61, respectively.
Superimposed on this general trend, one observes
some subtle features that can be visualized in a plot
of $\mathcal{R}^\alpha \, \phi(E)$ versus $\mathcal{R})$,
where the flux is multiplied by a factor that grows rapidly with
rigidity (with an exponent $\alpha$ of order 2.6--2.7)
that compensates for the rapid decrease of the fluxes.
In such a representation the proton and helium spectra
appear, in first approximation, to be formed by three main structures,
that  have often been referred to in the literature
(using a potentially misleading expression) as ``bumps''.

In both spectra the ``top'' of the bumps
(where the spectra undergo a softening)
are centered at approximately equal rigidities of order 30~GV, 15~TV and 3~PV.
The bumps are separated by two ``valleys'' (where the spectra undergo a hardening).
The lower--rigidity hardening is located at approximately the same rigidity
($\mathcal{R} \approx 0.5$~TV) for both protons and helium nuclei.
However, the second hardening is observed at different rigidities for protons
($\mathcal{R} \approx 100$~TV) and for helium nuclei
($\mathcal{R} \approx 500$~TV).
This corresponds to the fact that the ``middle bump'' of the helium spectrum is significantly
broader than the corresponding proton bump.
In fact, in this work
we fitted the helium spectrum in the relevant
range as a sequence of two softenings (centered at 15 and 100~TV).

The origin of the lowest rigidity ``bump'' is well understood.
It is formed because, for energies below a few tens of GeV, the CR fluxes are
strongly suppressed by solar modulations effects.
The interpretation of the other spectral features is still controversial.

Several authors \cite{Tomassetti2012,Blasi2012,AloisioBlasi2013}
have modeled the spectral hardening observed at $\mathcal{R} \simeq 0.5$~TV as
a propagation effect, associated with a change in the properties
of turbulence in interstellar space. This hypothesis receives support
from studies of the break at the same rigidities in the 
secondary--to--primary ratios (such as B/C) \cite{Genolini2017,EvoliBlasiMorlino2018}.
The other 
spectral features at higher energies  have generally  been attributed to
the properties of the CR sources, assuming that the effect of propagation,
can be described by a simple, featureless power--law.

In the multicomponent framework a
minimal model requires only two components, one dominant at lower energies 
and forming the ``middle bump'', and the other dominant at higher energies
and forming the next bump.
Such a minimal model is discussed 
by Aharonian and Zhang \cite{Aharonian:2026tzf}. 
In this model the proton spectrum is formed by 
a lower--energy component with slope $\Gamma_1^{p} \simeq 2.73$ 
that extends up to a maximum energy of order $E_1^{p} \simeq 65$~TeV,
where the flux has a (super)--exponential cutoff, and
a second harder component with slope $\Gamma_2^{p} \simeq 2.39$
that extends up to the cutoff energy $E_2^p \simeq 6.2$~PeV.
This combination of components allows to describe successfully
the three main features of the spectrum for $E \gtrsim 1$~TeV.
The lower--energy component is dominant in the range of the ``middle bump'', 
with its cutoff shaping the softening centered around 15~TeV.
The spectral hardening at $E \simeq 100$~TeV
corresponds to the transition energy where the contribution
of the second component becomes dominant.
The cutoff of the second component forms the spectral break
that corresponds to the ``knee''.

The modeling of the helium spectrum is similar.
It is also formed by two components,
with the lower--energy one having slope
$\Gamma_1^{\rm He} \simeq 2.57$ and cutoff energy $E_1^{\rm He} \simeq 450$~TeV,
and a second component constructed rescaling the corresponding
proton component assuming a rigidity--dependent shape
(equal slope and a cutoff energy twice as large).
The two components combine to form the features observed in the  helium flux.

A problem with this model is that, while the higher--energy components
of the proton and helium spectra a have the same shape in rigidity
(suggesting they have the same origin),
the lower--energies components have very different shapes.
They not only have different slopes
($\Gamma_1^p \simeq 2.73$ and $\Gamma_1^{\rm He} \simeq 2.39$),
but also have cutoffs energies (65 and 460~TeV)
that differ by a factor of order seven,
much larger that the factor of two predicted by rigidity--dependence.
A large difference in the values of the proton and helium cutoff energies 
is imposed by the fact (already discussed above) that the shapes in rigidity of the
``middle bumps'' in the two spectra differ,
with the helium bump being significantly broader.

These differences in the shapes of the proton and helium spectra
appears to contradict the hypothesis that they are produced
by the same sources.
However, the authors of  \cite{Aharonian:2026tzf} argue that
proton and helium spectra of different shapes
can be produced by sources in the same class
(proposing supernova explosions as the best candidate for such a class)
because the two nuclei are accelerated in different environments
of the sources. Nevertheless, an acceleration  model that incorporates these ideas
has not yet been constructed.

Other authors \cite{DAMPE:2025opn,Yuan:2025xqt,Li:2026uox}
have argued that to describe the proton and helium observations
in the  multicomponent framework  it is necessary to include
the contribution of a ``local source''. In these papers
the local source is assumed to be point--like and
to produce instantaneously CR spectra  with 
the ``standard'' shape  of Eq.(\ref{eq:q-standard})
(for each particle type a power--law with a  high--energy cutoff  proportional
to its  electric charge). The effects of propagation
are calculated using in a diffusion model and
taking into account the distance and age of the source
(that are considered free parameters).

The role of the local source is to form the ``middle bump'' in the
proton and helium spectra, while two other components, formed by the ensemble
of the accelerators in two  separate classes, are dominant at lower and higher energies.

The ``local source'' explanation is the favored one in the
paper of the DAMPE collaboration \cite{DAMPE:2025opn} that reported 
measurements of the spectra of five different elements
(proton, helium, carbon, oxygen and iron)  extending to the multi--TeV energy range.
The DAMPE paper claims that the observations establish that
all these five spectra contain a softening feature with a universal shape in rigidity,
centered at $\mathcal{R} \approx 15$~TV, and discuss an
interpretation of the observations where the 
spectra are modeled by the combination of a smooth
``background'' component   with  the contribution from one burst--like
``local'' source.
This source releases proton and helium spectra that
have power--law form with different slopes (2.48 and 2.38 respectively) and exponential
cutoffs at the same rigidity.
The distance and age of this ``local''source are estimated to be consistent with
those of the Geminga pulsar ($r \simeq 0.25$~kpc and $t \simeq 0.24$~Myr), and 
a possible anisotropy signature consistent with this association is also discussed
\cite{DAMPE:2025opn}).

The discussion in the DAMPE paper 
is limited to the range  $\mathcal{R} \lesssim 100$~TV and does not take into account
higher--energy measurements from EAS detectors.
As discussed above, the inclusion of higher--energy data shows that 
the features in the proton and helium spectra are not
rigidity dependent, and therefore the DAMPE analysis should be revised.

Other authors \cite{Yuan:2025xqt,Li:2026uox} have recently
published models for the proton and helium fluxes in a broader energy range
that extend to PeV energies and take into account the LHAASO measurements.
These authors model the CR spectra as formed by three components:
two ``background'' components associated to two differen classes of accelerators
and one local source. Following the approach of GST2013 \cite{Gaisser:2013bla}
the spectra of all three components are power--laws with exponential cutoff and
form ``Peter's cycles'' with cutoffs at the same rigidity.
The ``middle bump'' around $\mathcal{R} \sim 15$~TV is mainly due to the local source.
The two ``background'' components are dominant at lower and higher rigidities,
with the cutoffs of the higher--rigidity component associated to
the softenings in the PeV range.

In the work of Yuan \cite{Yuan:2025xqt} it is noted that the shapes
of the proton an helium spectra in the rigidity range
of the ``middle bump'' differ, and therefore  that a model that
describes this structure as  due  to the contributions of 
one local source with rigidity--dependent cutoffs
results in a poor fit to the data.
Always assuming that all sources  produce  spectra
with rigidity dependent cutoffs,
Yuan suggests that it could be necessary
to introduce not one, but two local sources. 
A proton rich source ($p$/He ratio of order 10) with cutoffs at $\mathcal{R} \simeq 15$~TV
and a helium rich source ($p$/He ratio of order 0.1)
with cutoffs at a four times higher rigidity ($\mathcal{R} \simeq 60$~TV).

The possibility that one (or perhaps more than one)
``local'' source gives a significant contribution to the CR fluxes
cannot be excluded. A model of the frequency and spatial distribution of
the Galactic CR sources allows to estimate the probability
that the contribution of local sources is significant, but the existence of
such contribution can only be established from observations.
However, it should be noted that in the models discussed  above 
the spectral shape of the local source
is inferred fitting the data and it differs from
the shapes of the ``background'' components.
Therefore the local source belongs to a distinct class of accelerators
with no other observed counterparts.

\subsection{Problems with multicomponent models}
Multicomponent models can reproduce reasonably well
the CR measurements. However, this success does not necessarily imply that
the models are correct, because they are described by many free parameters:
the slopes, energy cutoffs and normalizations of the different CR components.
Therefore some fundamental questions remain.
Are the model's key assumptions correct?
Do the model parameters accurately describe the properties of the sources in different classes? 

Three types of considerations suggest that exploring
alternative scenarios for the properties of the CR sources can be interesting. \\
(i) The construction of acceleration models that satisfy the requirements
associated to the multicomponent framework is difficult. \\
(ii) It is difficult to match the recent gamma--ray observations of Galactic sources,
that show that the sources have spectra with a wide variety of shapes,
with the prediction that the CR accelerators produce spectra
with only a few class--dependent shapes. \\
(ii) The features present in the proton and helium spectra 
(with the exception of the knee softening around $\mathcal{R} \simeq 3$~PV)
are subtle deviations (of order $\pm 20\%$) from a simple power--law.
This seems to require fine--tuning the relative sizes of the
contributions from different classes of sources,
because one would generally expect that the combination of components formed
by astrophysical sources of different nature (e.g. supernova explosions
or microquasars) and with very different cutoffs should result in
easily observable features in the combined spectrum.

\vspace{0.25cm}
Developing point (i) one can  observe that
the difficulty  in constructing a model of the accelerators  derives
from the fact that the multicomponent framework
requires that all sources belonging to the same class must produce 
spectra that, for each element, have in sufficiently good approximation
the same shape, that is the same slope and the same high--energy cutoff.
This ensures that the average of these spectra, 
(which determines the shape of the observable CR fluxes),
can also be described by a simple power--law with a sharp cutoff
[see Appendix~\ref{sec:combination-power-law} for a more quantitative discussion].
On the other hand, the shapes of spectra for different elements 
must have different slopes,
and perhaps  different rigidity cutoffs as well
(see the previous subsection for a discussion).
Therefore acceleration models, must identify the self--regulating mechanism that produces
nearly identical spectral shapes for each element in different sources,
while also explaining why different elements in each source
have spectra of different shapes.
The second requirement could 
be achieved by assuming that the acceleration of different particles occurs
in separate regions of the sources, as proposed by
Aharonian and Zhang \cite{Aharonian:2026tzf},
or alternatively by modifying the description of acceleration.
Malkov et al. \cite{Malkov:2011gb,Hanusch:2018bsk})
adopted the second approach, suggesting that
a revision of the standard DSA mechanism indicates that helium nuclei
are accelerated to a harder spectrum than protons. However,
this theory does not explain why,
for rigidities greater than approximately 100~TeV, the helium spectrum
becomes softer than the proton one.

\vspace{0.25cm}
Developing point (ii)  one can note that the problem  of connecting
cosmic--ray,  gamma--ray and neutrino emission is  a central problem
for high energy astrophysics.
Information about the spectra of freshly accelerated CR particles in astrophysical
sources can be obtained by observing gamma--ray and neutrino fluxes
produced by CR interactions inside or near their sources.
Space--based gamma--ray telescopes
\cite{EGRET:1999szp,Thompson:2008rw,AGILE:2009mvi,Fermi-LAT:2019yla,Fermi-LAT:2022byn,Ballet:2023qzs},
and ground--based gamma--ray telescopes
\cite{HESS:2018pbp,Acharyya:2023llt,LHAASO:2023rpg}
have observed a large number
of point--like or quasi--point--like sources belonging to several classes of
astrophysical objects. These results demonstrate that the acceleration of
electrically charged particles to relativistic energies is a common phenomenon
observed in many astrophysical objects,
even though it has not been possible to 
unambiguously establish which objects are the dominant CR sources.
One interesting and puzzling result is that 
the spectra emitted by these gamma--ray sources exhibit a very large variety of shapes \cite{Lipari:2025nta}.
It is unclear whether this result can be reconciled with the hypothesis
that the sources produce CR spectra with only a few class--dependent shapes.
Studying this problem is difficult because
the relationship between gamma--ray emission and the properties of 
CR populations inside an astrophysical source depends on several factors.
First, the gamma--ray emission can be both of hadronic or leptonic origin,
and depends on the spatial distributions of the relativistic particles 
and of their interactions targets, that can vary from source to source. 
Additionally, gamma--ray observations 
study a source's emission during a short observation period, 
but generally sources have much longer and more complex time evolutions.
For cosmic--ray production one is interested
in the emission integrated over the sources entire lifetimes.
Given these difficulties, it is not possible to reach firm conclusions, 
but reconciliating gamma--ray observations with the
assumption that the high--energy sources produce CR spectra with a few
simple shapes is challenging.

\vspace{0.25cm}
Developing point (iii)  one can  note again  that
the features in the proton and helium spectra below the knee
are deviations of order $\pm 20$\%  with respect to a  simple power--law
(See Figures~\ref{fig:proton_flux1} and~\ref{fig:helium_flux1}).
We argue here that this seems to require some ``fine--tuning''
of the relative size of the different components in a multicomponent framework.
For instance, in models that include a ``local'' source, its flux
must be about 10\% of the total flux formed by the ensemble of all
other sources in the Galaxy. The probability of living in a place 
and epoch where one encounters such a situation can be estimated
only with a large uncertainty  but it is low.

Models that  do not include local sources also
require some fine--tuning of the size of the components.
For example, in the Aharonian and Zhang model \cite{Aharonian:2026tzf}
the softening  of the proton spectrum at  $E \simeq 15$~TeV is described as the
effect of a super--exponential cutoff of a first  component
that is dominant at this and lower energies.
The effects of  this sharp cutoff are weak and observable
only with precision measurements, because they are
compensated by the growing contribution of a second, harder component
that is ``fine-tuned'' to produce a very smooth total flux.

An alternative explanation for the lack of prominence
of the features observed in the CR spectra.
is that the  spectral features are simply the result
of statistical fluctuations on a spectrum that is not
exactly, but only in good approximation of power--law form.
This explanation is perhaps not attractive because of a lack of predictive power,
but cannot be excluded a priori.

\vspace{0.25cm}
The arguments developed above are far from proving that the multicomponent
models are incorrect. However, they are sufficient to motivate
exploring different scenarios for the properties of cosmic ray emission from the
Galactic sources. 

\section{The many--shapes hypothesis}
\label{sec:many-shapes}
In this section we will present an alternative scenario 
to describe the production of Galactic  cosmic--rays where the 
sources produce spectra that are not limited to only a few shapes,
associated to a few different classes of sources,
but can instead have a very large variety of shapes.
Summing over the ensemble of all sources yields a cumulative spectrum
that, as required by the observations,
has a power--law form. However, 
this shape emerges as a statistical average and is not
present in the spectra of individual sources.

\subsection{An analogy: the AGN gamma--ray sources}
The idea that a power--law spectrum can be obtained by summing
spectra with a wide variety of shapes may seem surprising,
so it can be instructive to show how this idea can be applied
successfully to a quite different type of astrophysical observations:
gamma--ray emission from active galactic nucleus (AGN) sources.

Most of the gamma--ray sources observed by the Fermi telescope are active galactic nuclei
AGN of the blazar type.
The 4FLG--DR4 Fermi catalogue
\cite{Fermi-LAT:2019yla,Ballet:2023qzs}, based on 14 years of observations,
contains a total of 7194 sources. Of these 4002 
are associated with AGN objects of the blazar class.
For each source the catalogue provides a best fit which can take one of
three forms: a simple power--law (for 2425 objects),
a logparabola form (for 1571 objects) and 
a power--law with a super--exponential cutoff (for the remaining six objects).
These three types of fit account for 60\%, 40\% and 0.15\% of sources
with AGN associations, respectively.
Sources fitted with the
power--law form have on average fluxes approximately five times smaller
that those fitted with curved spectra, and it is likely
that most (if not all) of the spectra with a power--law form
do have a curvature that is undetected due to being below
the measurement's sensitivity threshold.

\begin{figure}[bt]

\begin{center}
\includegraphics[width=15cm]{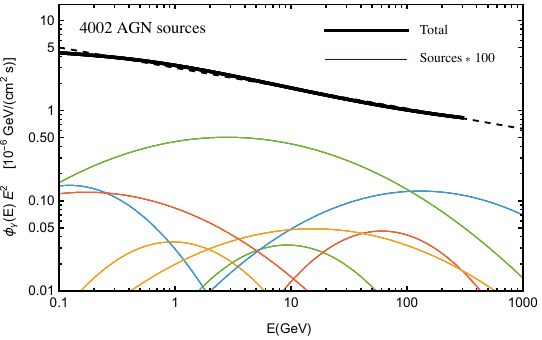}
\end{center}

\caption {\footnotesize
 \label{fig:AGN_cumulative}
 The thick line is the sum of
 the best fits to the 4002 sources in the Fermi 4FGL--DR4 catalog that
 are classified as AGN, plotted in the form
 $E^2 \, \phi_\gamma (E)$ versus $E$. The dashed line is a power--law approximation
 to the sum of the fits with a slope $\Gamma = 2.225$.
 For $E > 0.15$~GeV the power--law form approximation deviates from
 the exact result by less 7\%. The thin lines are the fits of some
 randomly chosen individual sources (of logparabola form) multiplied
 by a factor 100 for a better visualization.
}
\end{figure}

The sum of all 4002 best fits is shown in
Fig.\ref{fig:AGN_cumulative}, together with a
simple power--law function $\propto E^{-\Gamma_0}$ with a slope of $\Gamma_0 = 2.225$.
This form is a very good approximation
of the numerical result, for $E \ge 0.15$~GeV
the deviations are less than 7\%.

\begin{figure}[bt]

\begin{center}
\includegraphics[width=6cm]{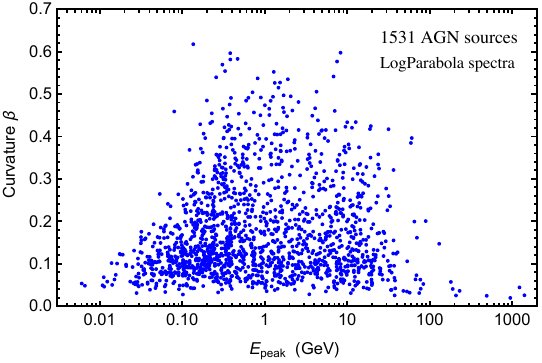}
~~
\includegraphics[width=6cm]{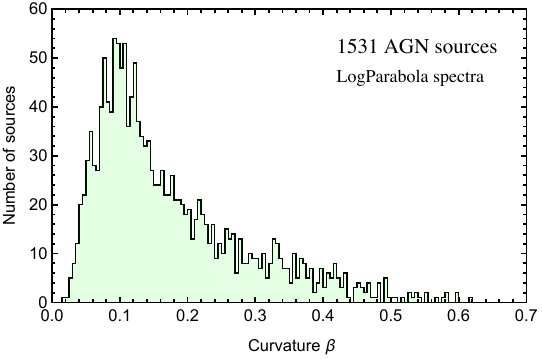}

\vspace{0.25cm}
\includegraphics[width=6cm]{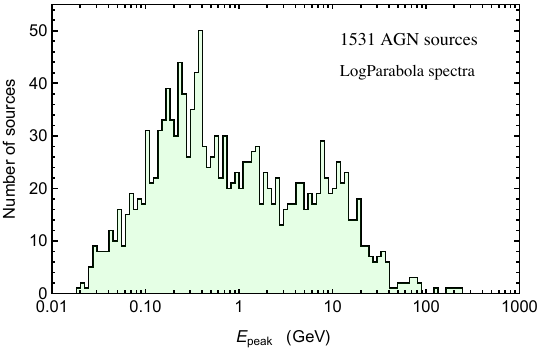}
~~
\includegraphics[width=6cm]{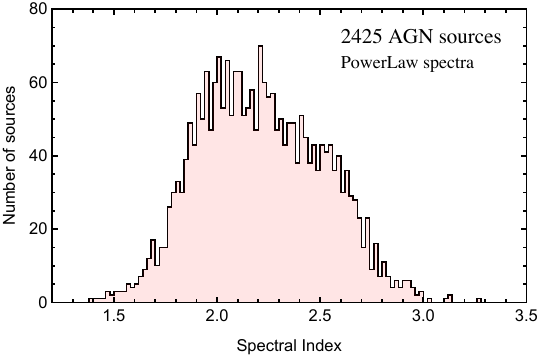}
\end{center}

\caption {\footnotesize
 \label{fig:AGN_parameters}
 Distribution of the fit parameters to
 the spectra of the 4002 sources in the FERMI 4FGL-DR4 catalog that are
 classified as AGN.
 In the catalog 1531 sources have best fits of logparabola form,
 2435 have best fits of power--law form, and 6 have fits of the
 cutoff form.
 For sources fitted with the logparabola form:
 the top--left panel shows a scatter plot of
 the shape parameters $\{E_{\rm peak}, \beta\}$,
 the top--right panel shows the distribution of
 the curvature parameter $\beta$, and the 
 bottom-left panel shows the distribution of
 the peak energy parameter $E_{\rm peak}$.
 The spectral index distribution for the sources
 fitted with the power--law form is shown
 in the bottom--right panel.
}
\end{figure}

Some examples of the spectra from individual AGN sources
(multiplied by a factor of 100) are also shown in 
Fig.\ref{fig:AGN_cumulative}. It is apparent that
the power--law form of the cumulative spectrum is obtained by summing contributions
that have a broad range of different shapes.
The shape of a power--law spectrum is determined by its
(energy independent) slope,
while the logparabola form has a slope that grows linearly
with energy
\begin{equation}
\Gamma (E) = 2 + 2 \beta \, \ln (E/E_{\rm peak})
\end{equation}
and is determined by two parameters: the energy $E_{\rm peak}$
where the slope takes the value of two, and the curvature $\beta$.
The distributions of the best parameters for
all fits of power--law and logparabola form 
are shown in Fig.~\ref{fig:AGN_parameters},
and are very broad.
For example, the values of $E_{\rm peak}$ extend over more than three orders of magnitude
(from 30~MeV to 100~GeV), and the slopes of the power--law fits
have values in the interval $(1.5 \lesssim \Gamma \lesssim 3$).

A similar discussion on spectra of the AGN sources in a previous version
of the Fermi--LAT catalog was presented in \cite{Lipari:2020szc}.


The interpretation of these results requires an in--depth discussion
that cannot be provided here. However, it is possible to
make some simple observations. The mechanism that
produces the gamma--ray observed in the AGN blazars by Fermi is not firmly established,
but it is likely to be the inverse Compton scattering
of relativistic $e^\mp$ interacting with the radiation fields present in the AGN jets.
While the target radiation fields 
in different sources are not expected to be identical, 
the large differences in the shape of the gamma--ray emission 
also indicates that the spectra of the $e^\mp$ that up--scatter
the photons can have very different shapes.

This suggests that the spectra of the CR produced by the Galactic
accelerators can also have different shapes,
and that the power--law form of their cumulative spectrum 
emerges only as an average.

\subsection{The power--law spectrum as a statistical average} 
\label{sec:logparabola}
In this section we will discuss how an ensemble of spectra of different
shapes can naturally combine to produce a cumulative spectrum of power--law form.
A crucial element is to assume that the spectral shape of each individual source 
is determined by one (source--dependent) characteristic energy.
For an explicit example, we will assume here that the spectra
have a logparabola form:
\begin{equation}
 q_j (E) = \sqrt{\frac{\beta_j}{\pi}} \;
\frac{\mathcal{E}_j} {(E_{*,j})^2} \;
\left ( \frac{E}{E_{*,j}} \right )^{-2 - \beta_j \, \ln (E/E_{*,j})} =
\frac{\mathcal{E}_j}{(E_{*,j})^2} 
 \; \overline{q}\left ( \frac{E}{E_{*,j}} , \beta_j \right ) 
~.
\label{eq:logparabola1}
\end{equation}
This spectrum is determined by three parameters: $\{\mathcal{E}_j, E_{*,j}, \beta_j\}$.
The parameter $\mathcal{E}_j$ gives the absolute normalisation of the spectrum
and has been chosen as the total energy of the cosmic--rays emitted by the source:
\begin{equation}
\int_0^\infty dE ~E \; q_j (E) = \mathcal{E}_j ~.
\end{equation}
The quantity $E_{*,j}$ is a characteristic energy that corresponds to the energy 
where the slope has a value of 2, 
or equivalently to the energy wat which the spectral energy distribution (SED)
$E^2 \, q_j(E)$ of the source has its maximum.
The adimensional quantity $\beta_j$, 
is the (energy independent) curvature, and is a second parameter that determines the spectral shape.
For the sake of simplicity, in the following we 
will assume that it has a unique value
(an average of its distribution can be performed at the end if desired).
With this simplification the spectral shape is determined by 
a single parameter, the peak energy $E_{*,j}$. Accordingly the spectrum has a scaling form
and can be expressed as a function of the ratio $E/E_{*,j}$.
See the last equality in Eq.~(\ref{eq:logparabola1}).

The cumulative spectrum can be obtained summing the contributions of all sources. 
Replacing the discrete sum with an integral over the source
parameters and introducing the (normalized) distribution
$dN_s/(d \mathcal {E} \, dE_*)$ 
and the frequency of the source events $T_s^{-1}$, the
cumulative spectrum can be calculated as:
\begin{equation}
 Q(E) = \frac{1}{T_s} \; \int \frac{dE_*}{(E_*)^2} \int d\mathcal{E} 
 \; \frac{dN_s} {d \mathcal {E} \, dE_* } ~ \mathcal{E} ~ \overline{q}
 \left (\frac{E}{E_*}\right )
 ~.
\label{eq:Q_int1}
\end{equation}
This expression can be rewritten as:
\begin{equation}
 Q(E) = \int \frac{dE_*}{(E_*)^2} \; \frac{dL_{CR}} {dE_*} ~ \overline{q}
 \left (\frac{E}{ E_*}\right ) ~,
\label{eq:Q_int2}
\end{equation}
where we have introduced the quantity:
\begin{equation}
 \frac{dL_{CR}} {dE_*} = 
 \frac{1}{T_s} \; \int d\mathcal{E} ~\mathcal{E} 
 \; \frac{dN_s} {d \mathcal {E} \, dE_*} 
\end{equation}
that is the (time averaged) power in CR emitted by all sources that have
peak energy $E_*$.

The crucial step is now to assume that the
distribution $dL_{CR}/dE_*$ has a power law form:
\begin{equation}
 \frac{dL_{CR}} {dE_*} = k_p \,\frac{ L_{\rm cr}} {E_0} \,
 \left ( \frac{E_*} {E_0} \right )^{-p}
\label{eq:sed_power}
\end{equation}
between some limits
$E^*_{\rm min}$ and $E^*_{\rm max}$.
In Eq.~(\ref{eq:sed_power})
$L_{\rm cr}$ is the total luminosity of the ensemble
of all CR sources,
$E_0$ is an arbitrary reference energy,
and $k_p$ is an
adimensional normalization constant:
\begin{equation}
 k_p = (p-1) \; \left (\frac{E^*_{\rm min}}{E_0} \right )^{p-1} \;
 \left [ 1 - \left (\frac{E^*_{\rm min}}{E^*_{\rm max}} \right )^{p-1} \right]^{-1} ~.
\end{equation}
With this assumption the integration over the peak energy $E_*$
can be performed analytically with the result:
\begin{equation}
 Q(E) = Q_0 ~
 \frac{L_{\rm cr}} {E_0^2} \; \left ( \frac{E}{E_0} \right )^{-(p+1)} 
\end{equation}
with $Q_0$ the adimensional quantity:
\begin{equation}
Q_0 = k_p \; \eta[E^*_{\rm min}/E, E^*_{\rm max}/E] \; e^{-(p-1)^2/(4 \beta)}
\end{equation}
The (also adimensional) quantity $\eta$ is a function
of the ratios $E/E^*_{\rm min}$ and $E/E^*_{\rm max}]$
that can be calculated with the integration:
\begin{equation}
 \eta[z_{\rm min}, z_{\rm max}] = e^{(p-1)^2/(4 \beta)} ~
 \int_{z_{\rm min}}^{z_{\rm max}} dz \; z^{-(p + \beta \, \ln z)} ~.
\end{equation}
For a broad range of $E$ $\eta$ is constant and equal to unity
going to zero only for very high energies ($E \gg E^*_{\rm max}$.

In summary, we assumed 
that the cosmic ray luminosity of the sources source is correlated
to the spectral shape of its emission,
so that sources that emit at higher energies, release less power,
and that this correlation follows a power--law form $L_{\rm cr} \propto (E_*)^{-p}$, 
where $E_*$ is the energy at which the source's spectral energy distribution (SED)
has its maximum.
Then the cumulative emission of the ensemble of all sources,
has a power--law form with slope $(p+1)$: $Q(E) \propto E^{-(p+1)}$.
This result was calculated explicitly above for logparabola spectra,
but is more generally valid for spectra whose shape is
determined by a characteristic energy.

According to this hypothesis, the power--law form of the CR spectra
does not emerge as the consequence of a ``universal'' mechanism that operates in all
sources of the same class. Rather, it is determined 
by the statistical properties of the ensemble of all sources in the class,
and is related to the fact that sources with harder spectra,
that produce higher--energy particles, are less frequent.

\subsection{The proton/helium ratio in multi--shapes models}
\label{sec:ratio}
A surprising result of the recent measurements
of the proton and helium spectra is not only that they 
contain ``features'' but also that in the rigidity ranges
where they are both well described by power--laws,
their slopes differ.
This result appears to be in conflict with the hypothesis that
both acceleration and propagation are rigidity--dependent.
If acceleration produces rigidity--dependent
source spectra of power--law form,
and propagation produces rigidity--dependent spectral distortions
also of power--law form, then the observable spectra are
power--laws with identical slopes for all particle types.

To explain this fact, multicomponent models
have assumed that for each class of accelerators the spectra of
the different elements are power--laws with rigidity--dependent
cutoffs but with different slopes. These slopes are estimated by
simply fitting the measurements without proposing a  physical
mechanism for the differences in slope.
As discussed above, a model that claims
to explain the slope difference observed in the lower energies range
has been proposed \cite{Malkov:2011gb,Hanusch:2018bsk}.
However, this model but does not address
the  fact that the slope difference $\Gamma_p - \Gamma_{\rm He}$
is different in different rigidity intervals
(changing also sign at $\mathcal{R} \simeq 100$~TV.

An interesting property of the many--shapes framework proposed in
this work, is that differences in slope between proton and helium nuclei
in a rigidity interval where both spectra have a power--law form
can naturally emerge even if both acceleration and
propagation are both rigidity--dependent.

As outlined above, if all accelerators 
produce spectra with the same universal shape, and therefore identical
slopes in all individual sources and for all particle types,
(and if propagation is also rigidity--dependent), 
then the proton to helium ratio is a constant.
This result does not require the sources to be identical,
or to produce identical $p$/He ratios. It only requires that
the shapes of the source spectra be identical
(and that the spectral distortions of
propagation be rigidity--dependent).

If different types of accelerators contribute to the observable
flux, and these accelerators produce source spectra of different shapes
and with different average $p$/He ratios, then the $p$/He ratio of the
observable fluxes will have a nontrivial rigidity--dependence.
As an illustration, if supernovae are the dominant CR sources at lower energies
producing softer spectra with a smaller average $p$/He ratio,
while microquasars are dominant at higher energies,
producing harder spectra with a larger average $p$/He ratio,
it is then obvious that the observable $p$/He ratio
will grow with energy (or rigidity). This is because
the fraction of the observable flux
due to the proton--rich microquasar component becomes larger.

However, in multicomponent models the rigidity range where the spectra
have a simple power--law form are those where a single class of sources
is dominant. This is because the combination of two (or more)
power--laws with different slopes forms a curved, gradually hardening spectrum,
and in this case the hypothesis of rigidity--dependent acceleration
implies equal slopes for all particle types.

In the many--shapes framework,
the hypothesis of rigidity--dependent acceleration
implies that 
the shapes of the rigidity spectra of protons and helium are identical
in each individual source. However, 
different sources produce spectra with different shapes and different
$p$/He ratios. 
The observable spectrum at different energies is formed by
different combinations of the spectra of individual sources,
(with spectra that are softer contributing more at lower energies,
and spectra that are harder spectra
contributing more at higher energies).
If the $p$/He ratio in individual
accelerators correlates to the hardness of their spectra,
then the $p$/He ratio of the observable fluxes will gradually change with rigidity.
In some sense, the many--shapes framework can be considered as a multicomponent
model with an infinite number of source classes.

Accordingly the decreasing $p$/He ratio observed by AMS
can be qualitatively explained by making the hypothesis 
that the sources producing the observable flux in this range 
are more helium--rich when their spectrum is harder.
Similarly, the observation that 
the helium spectrum is softer than the proton one
in the PeV energy range can be explained by assuming that
in this range harder sources become more proton--rich.
A physically motivated model to justify these hypothesis
needs to be developed.

\section{Summary and Outlook}
\label{sec:summary} 
For several decades, the measurements of the CR fluxes
suggested that over a broad energy range,
extending from a few tens of GeV,
where the effects of solar modulations
become negligible, to approximately 3~PeV,
where the all--particles spectrum
exhibits the break known as the ``knee'',
the cosmic ray spectrum had a simple power--law form
with a slope of approximately 2.7 for all nuclei.
During this period the scientific consensus was 
that the Galactic cosmic rays are produced by an ensemble of sources
that accelerate particles with a universal mechanism
commonly identified as diffuse shock acceleration (DSA), 
and generate power--law spectra with a unique exponent.
More precise measurements of the CR spectra have revealed
that, below the knee, the CR spectra do not have 
a simple power--law form but rather contain several features.
Additionally, in the rigidity intervals where the spectra are
well described by a power-law, the slopes of different elements differ.
These new observations require a revision of the CR models.

\vspace{0.2cm}
Essentially all works discussing these new observations
adopt the same ``multicompoment'' framework, and propose
extensions of the simple scheme outlined above, assuming
the validity of its most important ideas. The fundamental difference
is that the previous ``one-class of sources''
framework assumed the existence
of a single class of accelerators, producing CR spectra
with a universal shape. In contrast, the revised models 
assume that several
classes of astrophysical objects contribute to the the CR spectra,
and the sources in each class produces spectra
with the same class-dependent shape.
These spectral shapes all have the same general form: a power--law with
a high--energy exponential cutoff. However, the values of the
shape parameters (slopes and cutoff energies) are class--dependent.

Most of the models that adopt the ``multicomponent'' to framework
describe the CR spectra find it also necessary to include the
contribution of one (or perhaps two) ``local'' sources,
active in the recent past close to the solar system.
Also in this case the CR emission has the
standard shape of a power--law with exponential cutoff, 
however the effects of propagation depend on
the distance and age of the sources (that are treated as
free parameters in fits to the data).

\vspace{0.2cm}
In this work we advance the proposal that the Galactic
cosmic ray sources do not produce spectra with
characteristic shape, equal for all objects that belong to the same class,
but instead can have a wide range of possible shapes.
Since the observed cosmic--ray spectra have a power--law form,
it is necessary that the sum of the contributions of
all sources forms, at least in good approximation, a power--law.
This requirement might appear very difficult (if not impossible) to
achieve. How can a featureless power--law distribution emerge from
spectra with a large variety of shapes?
In fact this is possible, if two conditions are satisfied.
\begin{itemize}
\item The shape of the spectra of an individual source
is defined by a characteristic energy $E_*$ that is 
related to the average energy of the particles emitted by the source 
A possible choice  for this quantity (adopted in this work)
is the energy where the spectral energy distribution (SED) of
the CR emission has its maximum.
\item 
There is a correlation of power--law form
($L_{\rm cr} \propto E_*^{-p}$) between the luminosity 
(that is the power of the CR emission) $L_{\rm cr}$ of the sources 
and the characteristic energy $E_*$.
Such a correlation describes the fact that
the emission by soft (low $E_*$) sources
is larger than the emission by hard (high $E_*$) sources.
\end{itemize}
If these conditions are satisfied, then
the cumulative energy spectrum $Q(E)$ obtained summing the contribution
of all sources has also a power--law form: $Q(E) \propto E^{-(p+1)}$,
with a slope $p+1$ that is related to the slope $p$ of the correlation between
luminosity and characteristic energy.

\vspace{0.2cm}
In the many--shapes framework proposed in this work the slopes
of the CR spectra are  associated to the statistical
properties of the ensemble  of the sources.  This in  contrast to
frameworks where the slopes of the spectra can, at least in principle,
be calculated developing a model of the acceleration mechanisms.
This is can be seen as disappointing and as a limitation 
that makes the proposal unappealing.  On the other hand this  argument
cannot be used to falsify the proposal.
In fact, the mechanism outlined above for the production of a power--law
spectrum is actually verified in the case of the gamma--ray emission
from Blazars. The fourth catalog of gamma--ray sources of
the Fermi--LAT telescope, contains a large number (4002) of sources
associated to AGN of the Blazar type.
Summing the best fits (that have parameters 
in very wide intervals of values)  for all these sources
one obtains a cumulative spectrum that, in the energy range
between 100~MeV to more than 100~GeV (more than three orders of magnitude),
is well approximated, with deviations less than $\pm 7$\%,
by a simple power--law with a slope of 2.20.

\vspace{0.2cm}
The question remains of why  there is a correlation of power--law form
between the shape of a the CR emission and the luminosity of the spectra.
Unfortunately, also to this question  does not have at the moment
a simple answer.
In fact, power--laws appear widely in many fields.
One of the best examples is the distribution
of the energy released during earthquakes, which has been found
to obey the well known Gutenberg--Richter law \cite{gutenberg-richter}
that states that the logarithm of the earthquake frequency
is linearly related to their magnitude, which is proportional
to the logarithm of the released energy.
This law has been proposed many decades ago
on the basis of empirical observations,
but its origin has not yet been fully clarified.
There are many other examples of data sets described by
power--laws, for example the distributions of solar flares, moon craters,
forest fires, sizes of cities and many other quantities \cite{newman_powerlaws,clauset_powerlaws}.
The origin of these power--law distributions 
has been a topic of debate for over a century.
One possible explanation, for at least a number of them, is 
the intriguing concept of self--organized criticality,
by Bak {\it et al.} \cite{Bak:1987xua,Bak:1988zz}, who proposed it
to describe dynamical systems analogous to their paradigmatic sandpile model.
In any case, the possible existence 
of a correlation between the power output of the cosmic ray accelerators
and the spectral shape of their emission,
as proposed in this work, should be studied empirically.

\vspace{0.2cm}
A motivation for the introduction of the many--shapes framework is
to allow the  CR spectra to have larger deviations  from
a simple power--law form. 
In this paper we have shown that the more recently observed
features in the proton and helium spectra can be described as
deviations of order $\pm 20$\% on simple power--laws that extend
up the knee energy. Such small deviations ``look like'' statistical
fluctuations. To obtain spectra as ``smooth'' as those observed
for protons and helium, the shape paramaters (slopes and cutoffs)
and normalizations of the contributions of the different components
mut be ``fine--tuned'' to avoid the formation of more evident
spectral structures. This is of course possible, but
it appears as ``not--natural''.
On the other hand, if the energy distribution of the observed CR spectra emerge as
a statistical average from the combination of contributions
of different shape, one expects that it will not have 
a perfect power--law form, and because of statistical fluctuations,
one will observe some deviations. We speculate that at least
some of the features visible in the observed spectra could be
the result of these fluctuations.
Unfortunately, without a more detailed model of the sources
it is not possible to give a more quantitative estimate
of the size of the expected fluctuations to compare with the data.
A qualitative prediction is that more precise measurements
of the CR spectra will reveal more (and more subtle) features.

\vspace{0.2cm}
A second important motivation for the proposal of the many--shapes model
is that the current multicomponent framework has very stringent requirements
for the accelerators, that appear as very difficult to obtain.
In the multicomponent framework all sources in each class must produce
spectra with shapes that are nearly identical. These shapes
(for each element in the same class) are defined by a slope
and a high energy cutoff. Both parameters can take values in only a small
interval. This is because a spread in the values of the slopes results in
a curved cumulative spectrum (with the soft spectra of the individual sources
dominating a low energy,
and the hard spectra dominating at high energy).
Similarly a spread in the value of the high energy
cutoff results in a break more gradual than an exponential.

The prediction of a universal slope with a universal value
of two for acceleration in strong shocks was a very attractive feature
of the standard treatment of DSA.
Unfortunately a slope of two cannot be easily reconciled
with the observations, because the best estimates of the effects
of propagation do not produce a sufficient large softening 
to reproduce the data assuming such a hard source spectrum.

Multicomponent models require that acceleration 
produces spectra with slopes that are different for the different
classes of sources and, even more suprising, for the different elements.
For example in the two--classes model of Aharonian and Zhang
\cite{Aharonian:2026tzf}, the source spectra of the first component 
have slopes of $(2.73-\delta)$ and
$(2.57-\delta)$ for protons and helium nuclei, respectively,
while for the second component both spectra have a slope
of $(2.39 - \delta)$, with $\delta$ the exponent that describes
the propagation effects ($\propto \mathcal{R}^{-\delta}$).
The construction of models of the acceleration that produce spectra with
these slopes in all the individual objects of the same class
appears a very difficult task.

Similar considerations can be developed for the cutoff energies,
In this case the condition of equal values of the cutoffs in
different individual accelerators of the same class is perhas even more
diffcult to reproduce in a model, because the maximum energy can
in many cases depend also on the environment around the source.

Most of models developed in the multicomponent framework
assume that the cutoff energies scale with the electric charge of the
element considered, and therefore that in each class one has a single
rigidity cutoff. This is clearly a very attractive simplification.
It should however be noted that the assumption that, after correcting
for the differences in slope, the spectral
features (and therefore the cutoffs) of different elements
are rigidity--dependent is
inconsistent with the combination of the LHAASO and DAMPE data.
This problem is addressed by Aharonian and Zhang
\cite{Aharonian:2026tzf} simply accepting the fact that not only the slopes
but also the maximum acceleration rigidities 
of protons and helium nuclei
can be different in the same individual source.

\vspace{0.2cm}
An  attractive  result of the many--shape framework is that it
allows acceleration to be rigidity--depedent,
The observations that protons and helium nuclei have different slopes
is a problem that in the multicomponent models solves  simply 
assuming that acceleration, in contrast to what was previously
believed, is not rigidity--dependent.
In the many--shapes framework the difference in slopes
can be explained even assuming that all individual sources
produce proton and helium spectra with the same shape in rigidity.
This is because the slope of the observed spectra is related to
the correlation between the luminosity of the sources
and the shape of their spectra. This correlation can be different
for different elements. If for example harder sources are more rich in
helium nuclei with a correlation of power--law form:
$\langle q_{\rm He}/q_p \rangle \propto E_*^{\Delta p}$
(with $\langle q_{\rm He}/q_p \rangle$ the average helium to proton ratio,
that is independent from energy in each individual source)
then the observable helium spectrum will be harder than the proton
one by a factor
($\propto E^{-\Delta \Gamma})$ with $\Delta \Gamma = \Delta p$.

\vspace{0.2cm}
A comment is in order regarding the question of the nature
of the astrophysical objects that accelerate cosmic--rays in the Galaxy.
Multicomponent models are based on the assumption
that more than one class of sources contribute
to the CR spectra. This is an attractive idea, 
supported by the observations of the gamma--ray telescopes
that have detected fluxes from several types of astrophysical objects.
This idea can be also incorporated in the many--shapes framework.
In multicomponents models the contributions of different classes
of sources are dominant in different energy intervals, and
for example a particularly important question,
that remains the object of a lively debate,
is which class of sources form the ``Pevatrons'', that is the sources
that produces the CR particles in the knee region, with
microquasars or a subset of supernovae
considered the most likely candidates.
In the many--shapes framework it is possible that sources 
belonging to different classes contribute in the PeV energy range,
and it is also possible that different classes of sources
(such as stellar winds) could give significant contributions
to CR production at low energies.

\vspace{0.2cm}
A very important question, that clearly requires a critical and
extended discussion,  is if the idea proposed here,
that the Galactic accelerators produce  spectra of different shapes
is consistent with our understanding of  acceleration.,
and in particular with the theory of DSA 
that has been generally accepted for several decades.
We will argue here that the many--shapes framework
can be reconciled with DSA theory introducing some relatively
simple modifications.
In the ``textbook'' version of DSA theory a CR particle
acquires its energy in multiple steps,
repeatedly crossing the shock front.
In each step the energy increases by an amount proportional to its
initial value 
($\Delta E = \xi E$, with $\xi$ a small adimensional constant), and
$P_{\rm esc}$ is the probability of ``escaping'' from
the accelerator durung one step. Combining these equations,
one then obtains a power--law spectrum
of slope $\Gamma = 1 + P_{\rm esc}/\xi$.
This scheme can be modified to produced curved spectra.
For example, if the escape probability is
suppressed for low rigidity particles, or enhanced for high rigidity
this will reduce the spectrum both at low and high energies
and produce a convex spectrum (described in first
approximation by a logparabola spectrum).
For example, if the escape probability is
suppressed for low rigidity particles
this will reduce the spectrum at low energies,
and if it is suppressed for high rigidity particles
this will reduce the spectrum at high energies.
Combining these effects  it is possible to
obtain a convex spectrum, described in first
approximation by a logparabola form.
These effects are likely to be different from source to source
resulting in spectra of different shapes.
Developing these ideas it seems natural to
speculate that the linear relation between
$\log N$ and $\log E$ (with $N$ the number of
accelerated particles and $E$ their energy)
that describes the power--law prediction of standard DSA
is only the first approximation of a more complex
relation, with the quadratic relation of the logparabola form
the next order approximation.
These qualitative ideas should clearly be developed constructing
a reasonably realistic model. This task it postoponed to a future work.

\vspace{0.2cm}
The most promising method to test experimentally  the
ideas proposed in this work and the other frameworks proposed
to explain the cosmic ray measurements  is
the observation with gamma--rays and  neutrinos of the
Galactic high-energy sources.
These observations allow one to reconstruct (or constrain) the populations of
relativistic particles inside or near
high--energy sources. In fact one could argue that the wide
variety of spectral shapes observed by the gamma--ray telescopes
in sources of the same type is difficult to reconcile
with the hypothesis that these sources produce cosmic--ray spectra
of the same shape, as assumed by multicomponent models.
However, this conclusion is currently only tentative,
because large differences in the gamma--ray emission are 
also due to to the different environments in which sources are
found. Additionally, the predictions
discussed in this article are for CR emissions
integrated over the entire history of the sources
(assumed to be transients), whereas gamma--ray observations
are ``snapshots'' of the sources taken at various stages of
their evolution.

The relationship between CR production and the emission 
of gamma--rays and neutrinos is the core problem in high--energy astrophysics.
Current and future  observations of different types of
high--energy sources have the potential to clarify
where and how cosmic rays are accelerated.

\vspace{0.35cm}
\noindent{\bf Acknowledgments} \\
I gratefully acknowledge discussions with
Silvia Vernetto,
Carmelo Evoli,
Cao Zhen,
Dmitriy Khangoulian,
Felix Aharonian and
Bing Theodor Zhang


\appendix

\section{Combination of spectra of power--law form with exponential cutoff}
\label{sec:combination-power-law}

In this appendix we will assume that the CR sources produce spectra 
with the shape of a power--law an exponential cutoff,
defined by the slope $\Gamma$ and the cutoff energy $E_{\rm cut}$,
and discuss the shape of a linear combination of
these spectra if the values of $\Gamma$ and $E_{\rm cut}$ are not identical
but are distributed in an interval of values.

In the energy range where the effects of the cutoff are negligible
the spectra of an individual source (labeled by the index $j$) can be written as:
\begin{equation}
q_j (E) = q_{0,j} \, \left (\frac{E}{E_0} \right )^{-\Gamma_j} 
\end{equation}
(with $E_0$ an arbitrary reference energy).
The total spectrum is obtained summing the contributions of all sources:
\begin{equation}
 Q(E) = \sum q_j (E) \simeq \int d\Gamma
 \int dq_0 ~
 \frac{d^2 N_s} {dq_0 \, d\Gamma} \; q_0 ~.
 \left (\frac{E}{E_0} \right )^{-\Gamma}
\label{eq:q-cumul}
\end{equation}
In the second equality we have replaced
the discrete sum with integrations
over the spectral parameters ($q_0$ and $\Gamma$) 
introducing the distribution $d^2N_s/(dq_0 d\Gamma)$.

We will assume that at the (arbitrary) energy $E_0$ the distribution of the slopes
of the individual sources is a gaussian with average value $\Gamma_0$ and width $\sigma_\Gamma$.
This condition can be expressed assuming that the integral over 
the normalization parameter $q_0$ of the distribution $q_0 \, d^2N_s/(d\Gamma \, dq_0)$
has the form:
\begin{equation}
\frac{dQ}{d\Gamma} (\Gamma, E_0) = \int dq_0 ~
 \frac{d^2 N_s} {dq_0 \, d\Gamma} ~q_0 \simeq
 \frac{Q_0}{\sqrt{2 \pi} \, \sigma_\gamma}
 \; \exp \left [ -\frac{(\Gamma -\Gamma_0)^2}{2 \, \sigma_\gamma^2} \right ]~.
\end{equation}
Using this result in Eq.~(\ref{eq:q-cumul}), the integration 
over the slope $\Gamma$ can be performed analytically with the result:
\begin{equation}
Q(E) = Q_0 \; \left (\frac{E}{E_0} \right )^{-\Gamma_0 +\sigma_\gamma^2/2 \, \ln(E/E_0)} ~.
\end{equation}
This expression is a log--parabola spectrum with a
constant negative curvature $\beta = -\sigma_\Gamma^2/2$.
Accordingly the slope $\Gamma_Q (E)$ of the total spectrum is not a constant, but
decreases linearly with $\log E$:
\begin{equation}
\Gamma_Q (E) = \Gamma_0 - \sigma_\gamma^2 \, \ln(E/E_0) ~. 
\label{eq:gammaQ}
\end{equation}

It is interesting to note that the distribution of slope at the energy $E$
has the form:
\begin{equation}
\frac{dQ}{d\Gamma} (\Gamma, E) = 
 \frac{Q(E) }{\sqrt{2 \pi} \, \sigma_\gamma}
 \; \exp \left [ -\frac{(\Gamma -\langle \Gamma(E)\rangle)^2}{2 \, \sigma_\gamma^2} \right ]~.
\end{equation}
and is a gaussian with an energy dependent average 
\begin{equation}
 \langle \Gamma (E) \rangle = \langle \Gamma(E_0) \rangle
 - \sigma_\gamma^2 \, \ln(E/E_0)
\end{equation}
and the energy independent width $\sigma_\Gamma$.

The result of Eq.(\ref{eq:gammaQ}) implies that
a limit on the curvature of the spectrum ($|\beta| < |\beta^*|$)
can be interpreted as a limit on the spread of the slopes of the
individual spectra $\sigma_\gamma \lesssim \sqrt{2 |\beta^*|}$.

The models proposed to describe the CR spectra have a 
simple power--laws for several decades of energy, and this require
that the r.m.s. of the distribution of the slopes of the individual
spectra that form the component should have
a small spread of order $\sigma_\Gamma \lesssim 0.2$.

Similar considerations can be made for the case where the cutoff
energies of individual sources are not identical.

\clearpage

\bibliographystyle{apsrev4-2}
\bibliography{bibshapes}

\begin{table}
 \caption{\footnotesize Parameters of fits to the measurements of the
 cosmic ray proton and helium spectra by AMS02 \cite{AMS:2015tnn,AMS:2015azc},
 DAMPE \cite{DAMPE:2025opn} and LHAASO (QGJet analysis) \cite{LHAASO:2025byy,LHAASO:2025mlf}
 using the multi--break expression of Eq.~(\ref{eq:multi_break})
 with four breaks for protons and five breaks for helium nuclei.
 Energies are given in TeV, and the normalization $\phi_0$ (for the reference energy
 $E_0 = 1$~TeV) is in units (m$^{2}$~s~sr~GeV)$^{-1}$.
 \label{tab:par1}}
\begin{center} 
\renewcommand{\arraystretch}{1.25}

\begin{tabular}{ | c || c | c | c || c | c | c | }
\hline
~~~ & \multicolumn{3}{c||} {Proton Flux}
 & \multicolumn{3}{c|} {Helium Flux} \\
\hline
$b$ &
$\log_{10}[\mathcal{R}_b]$ & $\gamma_b$ & $w_b$ &
$\log_{10}[\mathcal{R}_b]$ & $\gamma_b$ & $w_b$ \\ 
\hline 
0 & -- & $2.75\pm 0.01$ & -- & -- & $2.68\pm 0.01$ & -- \\
1 & $2.81\pm 0.01$ & $2.57\pm 0.01$ & $0.45\pm 0.02$ & $2.89\pm 0.02$ & $2.42\pm 0.01$ & $0.55\pm 0.02$ \\
2 & $4.15\pm 0.01$ & $2.91\pm 0.01$ & $0.2\pm 0.1$ & $4.16\pm 0.01$ & $2.72\pm 0.01$ & $\lesssim 0.18$ \\
3 & $5.06\pm 0.01$ & $2.50\pm 0.01$ & $0.3^{+0.1}_{-0.2}$ & $5.03\pm 0.02$ & $2.91\pm 0.01$ & $0.23 \pm 0.03$ \\
4 & -- & -- & -- & $5.71\pm 0.05$ & $2.68\pm 0.03$ & $0.34\pm 0.10$ \\
5 & $6.50\pm 0.02$ & $3.4 \pm 0.1$ & $0.21 \pm 0.04$ & $6.56\pm 0.06$ & $3.7\pm 0.5$ & $\le 0.12$ \\
\hline
~~~
& \multicolumn{3}{c ||}{$\log_{10} [\phi_0$ (m$^{2}$~s~sr~GeV)$^{-1}]$} 
& \multicolumn{3}{c |}{$\log_{10} [\phi_0 $ (m$^{2}$~s~sr~GeV)$^{-1}]$} \\
\hline
& \multicolumn{3}{c ||}{$-2.20 \pm 0.01$} 
& \multicolumn{3}{c |}{$-3.12 \pm 0.01$} \\
\hline
\end{tabular}
\end{center}
\end{table}

\begin{table}
 \caption{\footnotesize Parameters of fits to the cosmic ray proton spectrum.
 Energies are given in TeV, and the normalizations
 $K$ in units (m$^{2}$~s~sr~GeV)$^{-1}$.
 \label{tab:par2}}
\begin{center} 
\renewcommand{\arraystretch}{1.25}

\begin{tabular}{ | c || c | c | c || c | c | c | }
\hline
~~~ & \multicolumn{3}{c||} {Proton Flux}
 & \multicolumn{3}{c|} {Helium Flux} \\
\hline
$b$ &
$\log_{10}[\mathcal{R}_b]$ & $\gamma_b$ & $w_b$ &
$\log_{10}[\mathcal{R}_b]$ & $\gamma_b$ & $w_b$ \\ 
\hline 
0 & -- & $2.80\pm 0.01$ & -- & -- & $2.73\pm 0.01$ & -- \\
1 & $2.83\pm 0.01$ & $2.56\pm 0.01$ & $0.20\pm 0.05$ & $2.89\pm 0.02$ & $2.42\pm 0.01$ & $0.55\pm 0.02$ \\
2 & $3.95\pm 0.01$ & $2.86\pm 0.01$ & $0.05\pm 0.1$ & $4.16\pm 0.01$ & $2.72\pm 0.01$ & $\lesssim 0.18$ \\
3 & $4.95\pm 0.01$ & $2.50\pm 0.01$ & $0.4^{+0.1}_{-0.2}$ & $5.03\pm 0.02$ & $2.91\pm 0.01$ & $0.23 \pm 0.03$ \\
4 & -- & -- & -- & $5.71\pm 0.05$ & $2.68\pm 0.03$ & $0.34\pm 0.10$ \\
5 & $6.50\pm 0.02$ & $3.4 \pm 0.1$ & $0.21 \pm 0.04$ & $6.56\pm 0.06$ & $3.7\pm 0.5$ & $\le 0.12$ \\
\hline
~~~
& \multicolumn{3}{c ||}{$\log_{10} [\phi_0$ (m$^{2}$~s~sr~GeV)$^{-1}]$} 
& \multicolumn{3}{c |}{$\log_{10} [\phi_0 $ (m$^{2}$~s~sr~GeV)$^{-1}]$} \\
\hline
& \multicolumn{3}{c ||}{$-2.20 \pm 0.01$} 
& \multicolumn{3}{c |}{$-2.86 \pm 0.01$} \\
\hline
\end{tabular}
\end{center}
\end{table}

\clearpage

`


`

\end{document}